\documentclass[sigconf,natbib=true]{acmart}

\usepackage{caption}
\usepackage{subcaption}
\usepackage{courier}  % DO NOT CHANGE THIS
\usepackage{algorithm}
\usepackage{algorithmic}
\usepackage{amsmath}
\usepackage{makecell}
\usepackage{tabularx}
\usepackage{threeparttable}
\usepackage{multirow}
\usepackage{booktabs}
\usepackage{marvosym}
\usepackage{bm}
\usepackage{mwe}
\usepackage{titletoc}
\usepackage{booktabs}
\usepackage{newfloat}
\usepackage{listings}
\usepackage{makecell}
\usepackage{threeparttable}
\usepackage{pdfpages}
\usepackage{import}
\usepackage[titletoc]{appendix}
\usepackage{multirow}
\usepackage{graphicx}
 \usepackage{enumitem}
 
\newcommand{\eg}{\emph{e.g.,}\xspace}

\newcommand{\ie}{\emph{i.e.,}\xspace}

\newcommand{\etal}{\emph{et al.}\xspace}

\newcommand{\wo}{\emph{w / o }\xspace}
\let\oldhat\hat

\renewcommand{\hat}[1]{\oldhat{\mathbf{#1}}}
\AtBeginDocument{%
  \providecommand\BibTeX{{%
    \normalfont B\kern-0.5em{\scshape i\kern-0.25em b}\kern-0.8em\TeX}}}

\copyrightyear{2023}
\acmYear{2023}
\setcopyright{acmlicensed}\acmConference[SIGIR '23]{Proceedings of the 46th International ACM SIGIR Conference on Research and Development in Information Retrieval}{July 23--27, 2023}{Taipei, Taiwan}
\acmBooktitle{Proceedings of the 46th International ACM SIGIR Conference on Research and Development in Information Retrieval (SIGIR '23), July 23--27, 2023, Taipei, Taiwan}
\acmPrice{15.00}
\acmDOI{10.1145/3539618.3591737}
\acmISBN{978-1-4503-9408-6/23/07}

\begin{document}

\title{Multi-view Hypergraph Contrastive Policy Learning for Conversational Recommendation}

\author{Sen Zhao}
\affiliation{%
  \institution{CCIIP Laboratory, Huazhong
University of Science and Technology
Joint Laboratory of HUST and Pingan
Property \& Casualty Research (HPL)}
  %\city{Wuhan}
  \country{China}}
\email{senzhao@hust.edu.cn}

\author{Wei Wei*}
\thanks{*Corresponding author.}
\affiliation{%
  \institution{CCIIP Laboratory, Huazhong
University of Science and Technology
Joint Laboratory of HUST and Pingan
Property \& Casualty Research (HPL)}
  %\city{Wuhan}
  \country{China}}
\email{weiw@hust.edu.cn}

\author{Xian-Ling Mao}
\affiliation{%
  \institution{Beijing Institute of Technology}
  %\city{Wuhan}
  \country{China}}
\email{maoxl@bit.edu.cn}

\author{Shuai Zhu}
\affiliation{%
  \institution{Ant Group}
  %\city{Wuhan}
  \country{China}}
\email{zs261988@antgroup.com}

\author{Minghui Yang}
\affiliation{%
  \institution{Ant Group}
  %\city{Wuhan}
  \country{China}}
\email{minghui.ymh@antgroup.com}

\author{Zujie Wen}
\affiliation{%
  \institution{Ant Group}
  %\city{Wuhan}
  \country{China}}
\email{zujie.wzj@antgroup.com}

\author{Dangyang Chen}
\affiliation{%
  \institution{Ping An Property \& Casualty
Insurance Company of China, Ltd}
  %\city{Wuhan}
  \country{China}}
\email{chendangyang273@pingan.com.cn}

\author{Feida Zhu}
\affiliation{%
  \institution{Singapore Management University}
  %\city{Wuhan}
  \country{Singapore}}
\email{fdzhu@smu.edu.sg}
\renewcommand{\shortauthors}{Sen Zhao et al.}
\begin{abstract}
  Conversational recommendation systems (CRS) aim to interactively acquire user preferences and accordingly recommend items to users. Accurately learning the dynamic user preferences is of crucial importance for CRS. Previous works learn the user preferences with pairwise relations from the interactive conversation and item knowledge, while largely ignoring the fact that factors for a relationship in CRS are multiplex. Specifically, the user likes/dislikes the items that satisfy some attributes (\emph{Like/Dislike view}). Moreover social influence is another important factor that affects user preference towards the item (\emph{Social view}), while is largely ignored by previous works in CRS. The user preferences from these three views are inherently different but also correlated as a whole. The user preferences from the same views should be more similar than that from different views. The user preferences from \emph{Like View} should be similar to \emph{Social View} while different from \emph{Dislike View}.
  To this end, we propose a novel model, namely \underline{M}ulti-view \underline{H}ypergraph \underline{C}ontrastive \underline{P}olicy \underline{L}earning (MHCPL). Specifically, MHCPL timely chooses useful social information according to the interactive history and builds a dynamic hypergraph with three types of multiplex relations from different views. The multiplex relations in each view are successively connected according to their generation order in the interactive conversation.
A hierarchical hypergraph neural network is proposed to learn user preferences by integrating information of the graphical and sequential structure from the dynamic hypergraph.
A cross-view contrastive learning module is proposed to maintain the inherent characteristics and the correlations of user preferences from different views. Extensive experiments conducted on benchmark datasets demonstrate that MHCPL outperforms the state-of-the-art methods.
\end{abstract}





 \maketitle
\section{Introduction}
% CRS MCR 简单介绍
 Recommendation systems \cite{zhao2022multi,wu2019session, wang2020disenhan,wei2023recommendation,ferrari2019we} are emerging as an efficient tool to help users find items of potential interest. They conventionally learn user preferences from their historical actions \cite{he2017neural,rendle2010factorization}, while hardly acquiring dynamic user preferences which often drift with time. To this end, conversational recommendation systems (CRS) \cite{lei2018sequicity} are proposed to dynamically acquire user preferences and accordingly make recommendations through interactive conversations. 
 \begin{figure}[ht]
    \centering
     \includegraphics[width=0.45\textwidth]{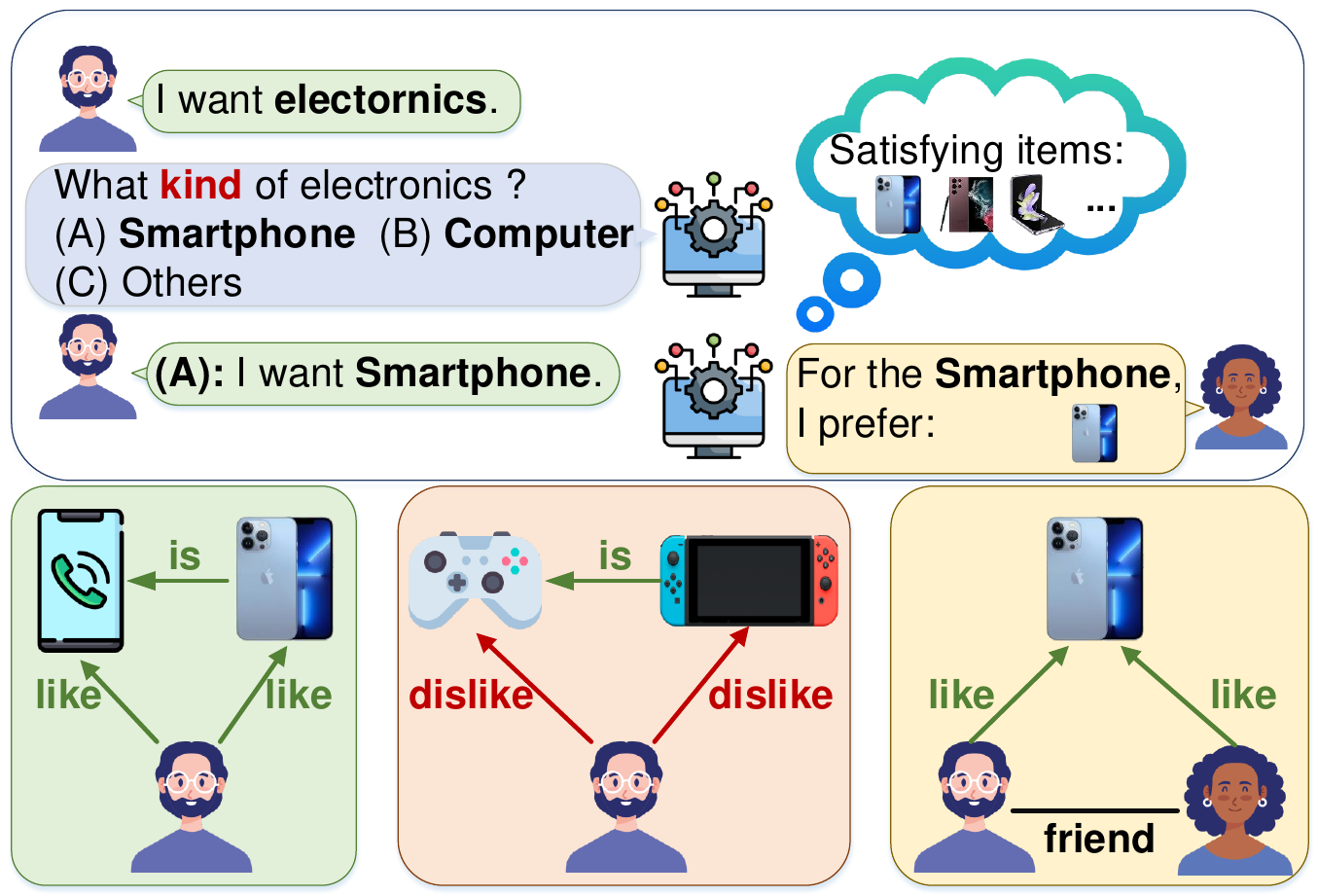}
    \caption{Common types of multiplex user relations from different views in the scenario of conversational recommendation.}
    \label{fig:motivation}
\end{figure}
 Different settings  \cite{christakopoulou2016towards,christakopoulou2018q,sun2018conversational} of CRS are explored and we focus on the Multi-Interest Multi-round Conversational Recommendation (MMCR) \cite{zhang2022multiple} in which users could accept multiple items and CRS needs to strategically ask multi-choice questions about user-preferred attributes and accordingly recommend items, reaching success in the limited turns. 

Learning the dynamic user preferences for the candidate attributes and items accurately is of crucial importance for CRS. CRM \cite{sun2018conversational} and EAR \cite{lei2020estimation} develop factorization-based methods to learn user preferences from pairwise interactions, but they fail to capture multi-hop information from the connectivity graph. SCPR \cite{lei2020interactive} learns user preferences by reasoning the path on the user-item-attribute graph.  Unicorn \cite{deng2021unified} and MCMIPL \cite{zhang2022multiple} further apply graph neural networks to 
 learn user preferences from the graph structure that captures rich correlations among different types of nodes (\ie user, attribute, and item). 
 Despite effectiveness, previous works learn user preferences with pairwise relations from the interactive conversation (\ie user-item and user-attribute relations) and the item knowledge (\ie item-attribute relations), while largely ignoring the fact that factors for a relationship in CRS are multiplex. For the example in Fig.\ref{fig:motivation}, the user dislikes Switch because of its attribute named "game console" rather than its other attributes like "electronics". Moreover, social influence is also an important factor that affects user preferences towards the item, since people with social connections will influence each other, leading to similar interests \cite{cialdini2004social, guo2015trustsvd}.
 However, in the field of CRS, social information is seldom explored.
 Inspired by the advantage of hypergraph \cite{feng2019hypergraph, xia2022hypergraph} in modeling the multiplex relations (\ie relations that connect more than two nodes), we investigate the potential of hypergraph modeling with the integration of interactive conversation, item knowledge, and social influence for learning dynamic user preferences in CRS.

Actually, it's non-trivial to build a hypergraph for learning dynamic user preferences in CRS, due to three challenges: 1) The first challenge is the dynamic filtering and utilizing of social information. The social information conventionally contains all the historical interactions of the user's friends, which could be noisy for the dynamic user preferences in the current conversation, since only friend preferences that satisfy the current conversation are helpful for dynamic user preferences learning.
For the example in Fig.\ref{fig:motivation}, only the friends' preferences for "smartphone" are helpful for learning the dynamic user preferences. 
2) The second challenge is hypergraph formulation. In the scenario of CRS (as illustrated in Fig.\ref{fig:motivation}), there mainly remain three multiplex relation patterns, that is, the user likes/dislikes the items that satisfy some attribute (\emph{Like/Dislike view}) and the user shares the preferences for items with some friend (\emph{Social view}). Each relation pattern corresponds to a kind of hyperedges, which are successively generated during the interactive conversation.
3) The third challenge is the aggregation of user preferences learned from different views, which might obscure the inherent characteristics of preference distributions from different views and the correlation between them. Specifically, user preferences from the same views should be more similar than user preferences from different views. And the user preferences from \emph{Like View} should be similar to \emph{Social View} while different from \emph{Dislike View}. Contrastive learning \cite{wu2021self, velickovic2019deep, hassani2020contrastive}, one successful self-supervised learning paradigm, which aims to learn discriminative representations by contrasting positive and negative samples, paves a way to maintain the inherent characteristics and the correlation of user preferences learned from different views.

To this end, we propose a novel hypergraph-based model, namely \underline{M}ulti-view \underline{H}ypergraph \underline{C}ontrastive \underline{P}olicy \underline{L}earning (MHCPL). Spe-cifically, MHCPL dynamically filters social information according to the interactive conversation and builds a dynamic multi-view hypergraph with three types of multiplex relations from different views: the user likes/dislikes the items that satisfy some attribute (\emph{Like/Dislike view}) and the user shares the preferences for items with some friend (\emph{Social view}). The multiplex relations in each view are successively connected according to their generation order in the interactive conversation. 
A hierarchical hypergraph neural network is proposed to learn user preferences by integrating information of the graphical and sequential structure from the dynamic hypergraph.
Furthermore, a cross-view contrastive learning module is proposed with two terms to maintain the inherent characteristics and the correlations of user preferences from different views.
%Furthermore, a contrastive learning module comparing user preferences learned from different views is developed to help integrate user preferences from different views. 
Extensive experiments conducted on Yelp
and LastFM demonstrate that MHCPL
outperforms the state-of-the-art methods.

\textbf{Our contributions} of this work are summarized as follows:
\begin{itemize}
    \item \textbf{General Aspects:} We emphasize the importance of multiplex relations and investigate three views to integrate interactive conversation, item knowledge, and social influence for dynamic user preference learning in CRS.
    
    \item \textbf{Novel Methodologies:} We propose the model MHCPL to timely filters social information according to the interactive conversation and learns dynamic user preferences with three types of multiplex relations from different views. Moreover, a cross-view contrastive learning module is proposed to maintain the inherent characteristics and the correlations of user preferences from different views.
    
    \item \textbf{Multifaceted Experiments:} We conduct extensive experiments on two benchmark datasets. The results demonstrate the advantage of our MHCPL in better dynamic user preference learning, which shows the effectiveness of our MHCPL for conversational recommendation.
\end{itemize}

\section{Related Works}
\subsection{Conversational Recommendation}
 Conversational recommendation systems (CRS) \cite{lei2018sequicity, priyogi2019preference, xie2021comparison, zhou2020improving,zhao2023towards} aim to communicate with the user and recommend items based on the attributes explicitly asked during the conversation. Due to its ability to dynamically get the user's feedback, CRS has become an effective solution for capturing dynamic user preferences and solving the explainability problem. Various efforts have been conducted to explore the challenges in CRS which can mainly be categorized into two tasks: dialogue-biased CRS studies the dialogue understanding and generation \cite{li2018towards,chen2019towards,kang2020recommendation,liu2020towards}, and recommendation-biased CRS explores the strategy to consult and recommend \cite{christakopoulou2016towards,christakopoulou2018q,sun2018conversational,lei2020estimation}. This work focuses on the recommendation-biased CRS.

Early works on the recommendation-biased CRS \cite{christakopoulou2016towards, christakopoulou2018q, sun2018conversational} only consider the conversational recommendation under simplified settings. For example, Christakopoulou  \etal \cite{christakopoulou2016towards} consider the situation that CRS only needs to recommend without asking the user about his/her preferred attributes. The Q\&A work \cite{christakopoulou2018q} proposes to explore the situation that CRS jointly asks attributes and recommends items, but restricts the conversational recommendation to two turns: one to ask attributes and the other to recommend items. To explore a more realistic scenario of the recommendation-biased CRS, further efforts \cite{lei2020estimation,lei2020interactive} based on the reinforcement learning (RL) are conducted to explore the problem of multi-round conversational recommendation (MCR) which aims to strategically ask users binary questions towards attributes and recommend items in multiple rounds, achieving success in the limited turns. Zhang \etal \cite{zhang2022multiple} further explore the setting of multi-interest MCR (MMCR) where users have multiple interests in attribute combinations and allows CRS to ask multi-choice questions towards the user-preferred attributes.

 The main challenge of MCR is how to dynamically learn user preferences, and accordingly choose actions that satisfy user preferences. CRM \cite{sun2018conversational} and EAR \cite{lei2020estimation} learn user preferences with a factorization-based method under the pairwise Bayesian Personalized Ranking (BPR) framework \cite{rendle2009bpr}. SCPR \cite{lei2020interactive} learns user preferences by reasoning the path on the user-item-attribute graph and strictly chooses actions on the path. Unicorn \cite{deng2021unified} builds a weighted graph to model the dynamic relationship between the user and the candidate action space and proposes a graph-based Markov Decision Process (MDP) environment to learn dynamic user preferences and choose actions from the candidate action space. MCMIPL \cite{zhang2022multiple} further considers the multiple interests of the user and develops a multi-interest policy learning module that combines the graph-based MDP with the multi-attention mechanism. Despite effectiveness, previous works model user preferences with binary relations, while hardly capturing the multiplex relations and ignore the influence of social relations on user preferences which are important in modeling dynamic user preferences.

\subsection{Social Recommendation}
Social recommendation \cite{kautz1997referral,guo2015trustsvd,jiang2014scalable} aims to exploit social relations to enhance the recommender system. According to the social science theories \cite{anagnostopoulos2008influence,bond201261,mcpherson2001birds}, user decisions are influenced by their social relations, leading to similar preferences among social neighbors. Following this assumption, SoRec \cite{ma2009trust} jointly factorizes the user-item matrix and the user-user social relation matrix by sharing the same user preference latent factor. STE \cite{ma2009learning} learns user preferences by linearly combing the preference latent factor of the user and his/her social neighbors. SocialMF \cite{jamali2009trustwalker} forces the user preference latent factor to be similar to that of his/her social neighbors by adding regularization to the user-item matrix factorization. These works only leverage first-order social neighbors for recommendation and ignore the fact that the social influence could diffuse recursively through social networks.

To model the high-order social influence, graph neural networks (GNNs) \cite{kipf2016semi} are introduced to social recommendation due to their superiority in learning the graph structure. GraphRec \cite{fan2019graph} applies GNNs to capture the heterogeneous graph information from the user-item interactions and social relations. DiffNet \cite{wu2019neural} and its extension DiffNet++ \cite{wu2020diffnet++}  develop a layer-wise influence propagation structure to model the recursive social diffusion in social recommendation. These works model user preferences with pairwise relations and fail to capture the complex multiplex user relation patterns (\ie user-friend-item). MHCN \cite{yu2021self} constructs hypergraphs by unifying nodes that form specific triangular relations and applies hypergraph neural network \cite{feng2019hypergraph, xia2022hypergraph} to model user preferences with hypergraphs. Despite effectiveness, previous works treat social relations as static information, while ignoring the dynamic characteristic of user preferences and failing to dynamically choose helpful social information for the learning of user preferences.

\section{DEFINITION AND PRELIMINARY}\label{sec:def}
In this section, we formulate the problem of multi-interest Multi-round Conversational Recommendation (MMCR) \cite{zhang2022multiple}.

Specifically, we define the set of items $\mathcal{V}$, attributes $\mathcal{P}$, and attributes types $\mathcal{C}$. Each item $v \in \mathcal{V}$ is associated with a set of attributes  $\mathcal{P}_v \subseteq \mathcal{P}$ and each attribute $p$ has its corresponding type $c_p \in \mathcal{C}$. In each episode, there exists an item set $\mathcal{V}_u$ that is acceptable for the user. Then CRS screens out candidate items $\mathcal{V}_{cand} \subseteq \mathcal{V}$ that contains the user-preferred attribute $p_0$ and candidate attributes $\mathcal{P}_{cand} \subseteq \mathcal{P}$ that are associated to the candidate items. Then in each turn $t$ ($t= 1, 2, \cdots, T$; $T$ is the max turn of the session), the CRS can either \emph{ask} $K_p$ attribute $\tilde{\mathcal{P}}_c \in \mathcal{P}_{cand}$ corresponding to the same attribute type $c$, or \emph{recommend} $K_v$ items  $\tilde{\mathcal{V}} \in \mathcal{V}_{cand}$:
\begin{itemize}
    \item If the CRS chooses to \emph{ask}, the user gives feedback according to whether $\mathcal{P}^{*}_c$ is associated with one of the items in the target item set $\mathcal{V}_u$.
    \item If the CRS chooses to \emph{recommend}, the user chooses to accept or not according to whether one of the items in the target item set $\mathcal{V}_u$ is listed in the recommended items $\tilde{\mathcal{V}}$.
\end{itemize}
The session of MMCR terminates if the user accepts the recommended items or leaves impatiently when the max turn accesses.

% \section{METHODOLOGIES}
% motivation of our technologies

\section{Framework}
In this section, we propose a novel Multi-view Hypergraph Contrastive Policy Learning (MHCPL) illustrated in Figure \ref{fig:model} that learns user preferences from the hypergraph integrating the interactive conversation, item knowledge and social information, and accordingly chooses actions. The Markov Decision Process (MDP) \cite{sutton2018reinforcement} formulation of our framework contains four components: multi-view user preference modeling, action, transition, and reward.

\subsection{Multi-view User Preference Modeling}
We first encode the state $s_t$, which contains the interactive conversation information $\mathcal{I}_u$ between the user and CRS, and the social information $\mathcal{F}_u$ that helps learn user preferences:
    \begin{equation}
    s_t=[\mathcal{I}^{(t)}_u, \mathcal{F}^{(t)}_u],
    \label{eq:state}
    \end{equation}
where $\mathcal{I}^{(t)}_u=[\mathcal{P}^{(t)}_{acc}, \mathcal{P}^{(t)}_{rej}, \mathcal{V}^{(t)}_{rej}, \mathcal{P}^{(t)}_{cand}, \mathcal{V}^{(t)}_{cand}]$ records the interactive history, and $\mathcal{F}^{(t)}_u$ denotes user's friends who have preferred items satisfying the interactive history $\mathcal{I}^{(t)}_u$, which is updated by:
 \begin{equation}
    \begin{aligned}
    \mathcal{F}_{u}^{(t)}=\{f \mid f \in \mathcal{F}_{u} & \text { and } \mathcal{V}^{(t)}_{f} \neq \emptyset\},
    \end{aligned}
    \label{eq:action}
    \end{equation}
where $\mathcal{F}_{u}$ denotes the friends of the user,  $\mathcal{V}^{(t)}_{f}=\mathcal{V}_{f} \cap \mathcal{V}_{cand}^{(t)}$ indicates the set of items that are acceptable for the friend $f$ and satisfy the interactive history. To this end, we build a dynamic hypergraph that integrates the interactive conversation, item knowledge, and social information to learn the user preference representation. Moreover, we develop a hypergraph-based state encoder to learn user preferences with multiplex relations from different views.
\subsection{Action}    
 According to the state $s_t$, the CRS agent chooses an action $a_t$ from the action space $\mathcal{A}_t$. The action space $\mathcal{A}_t$ contains candidate attributes $\mathcal{P}^{(t)}_{cand}$ and candidate items $\mathcal{V}^{(t)}_{cand}$, which are updated by:
    \begin{equation}
    %\mathcal{V}^{(t)}_{cand}=\mathcal{V}_{\mathcal{P}^{(t)}_{acc}}\setminus\mathcal{V}^{(t)}_{rej}
    \begin{aligned}
    \mathcal{V}_{\text {cand }}^{(t)}=\left\{v \mid v \in \mathcal{V}_{p_{0}}-\mathcal{V}_{r e j}^{(t)}\right.& \text { and } \mathcal{P}_{v} \cap \mathcal{P}_{acc}^{(t)} \neq \emptyset \\
    &\text { and } \left.\mathcal{P}_{v} \cap \mathcal{P}_{r e j}^{(t)}=\emptyset\right\},
    \end{aligned}
    \label{eq:action}
    \end{equation}
    
    \begin{equation}
    \mathcal{P}^{(t)}_{cand}=\mathcal{P}_{\mathcal{V}^{(t)}_{cand}}-\mathcal{P}^{(t)}_{acc}\cup\mathcal{P}^{(t)}_{rej},
    \label{eq:action2}
    \end{equation}
    where $\mathcal{V}_{p_{0}}$ denotes the items that satisfy the initial attribute $p_0$ of the user and $\mathcal{P}_{\mathcal{V}^{(t)}_{cand}}$ indicates attributes that belong to at least one of the candidate items $\mathcal{V}^{(t)}_{cand}$. When the CRS agent chooses to recommend, the agent chooses the top-K items $\tilde{\mathcal{V}}^{(t)}$ from $\mathcal{A}_t$. If the CRS agent decides to consult, the agent chooses $K_a$ attributes $\tilde{\mathcal{P}}^{(t)}_c$ that belong to the same attribute type $c$ from $\mathcal{A}_t$.
    
\subsection{Transition}  
After the CRS agent chooses the action $a_t$, the state $s_t$ will transition to the next state $s_{t+1}$. Specifically, if the agent chooses to consult, the attribute the user accepts and rejects in the current turn can be defined as $\mathcal{P}^{(t)}_{cur\_acc}$ and $\mathcal{P}^{(t)}_{cur\_rej}$. Then the state is updated by $\mathcal{P}^{(t+1)}_{acc}=\mathcal{P}^{(t)}_{acc}\cup\mathcal{P}^{(t)}_{cur\_acc}$ and $\mathcal{P}^{(t+1)}_{rej}=\mathcal{P}^{(t)}_{rej}\cup\mathcal{P}^{(t)}_{cur\_rej}$. When the agent chooses to recommend items $\tilde{\mathcal{ V}}^{(t)}$ and the user rejects all the items, the state is updated by $\mathcal{V}^{(t+1)}_{rej}=\mathcal{V}^{(t)}_{rej}\cup\tilde{\mathcal{ V}}^{(t)}$. Otherwise, this session ends with success. 
    
\subsection{Reward}
In this work, we design five kinds of rewards following previous works \cite{zhang2022multiple}: (1) $r_{rec\_suc}$, a strong reward when recommending successfully; (2) $r_{rec\_fail}$, a weak penalty when the user rejects the recommended items; (3) $r_{ask\_suc}$, a weak reward when the user accepts the asked attributes; (4) $r_{ask\_fail}$, a weak penalty when the user rejects the asked attributes; (5) $r_{quit}$, a strong penalty when the session quits without success. The reward on the multi-choice question is designed as $r_t=\sum_{\mathcal{P}_{cur_{-}acc}^{(t)}}r_{ask_{-}suc}+\sum_{\mathcal{P}_{cur_rej}^{(t)}}r_{ask_{-}rej}$.

\section{Multi-view Hypergraph Contrastive Policy Learning}
In this section, we detail the design of the Multi-view Hypergraph Contrastive Policy Learning (MHCPL) \ref{fig:model}. To model the dynamic user preferences, we build a hypergraph with three types of multiplex relations from different views to integrate information from the interactive conversation, item knowledge, and social information. To comprehensively learn user preferences, we develop a hypergraph-based state encoder, that captures the graphical and sequential structure in the dynamic hypergraph and propose a cross-view contrastive learning module to maintain the inherent characteristics and the correlation of user preferences from different views. Moreover, we develop an action decision policy to decide the next action based on the learned dynamic user preferences.

\subsection{Multi-view Hypergraph Construction}
\begin{figure*}[t]
    \centering
    \includegraphics[width=0.9\textwidth]{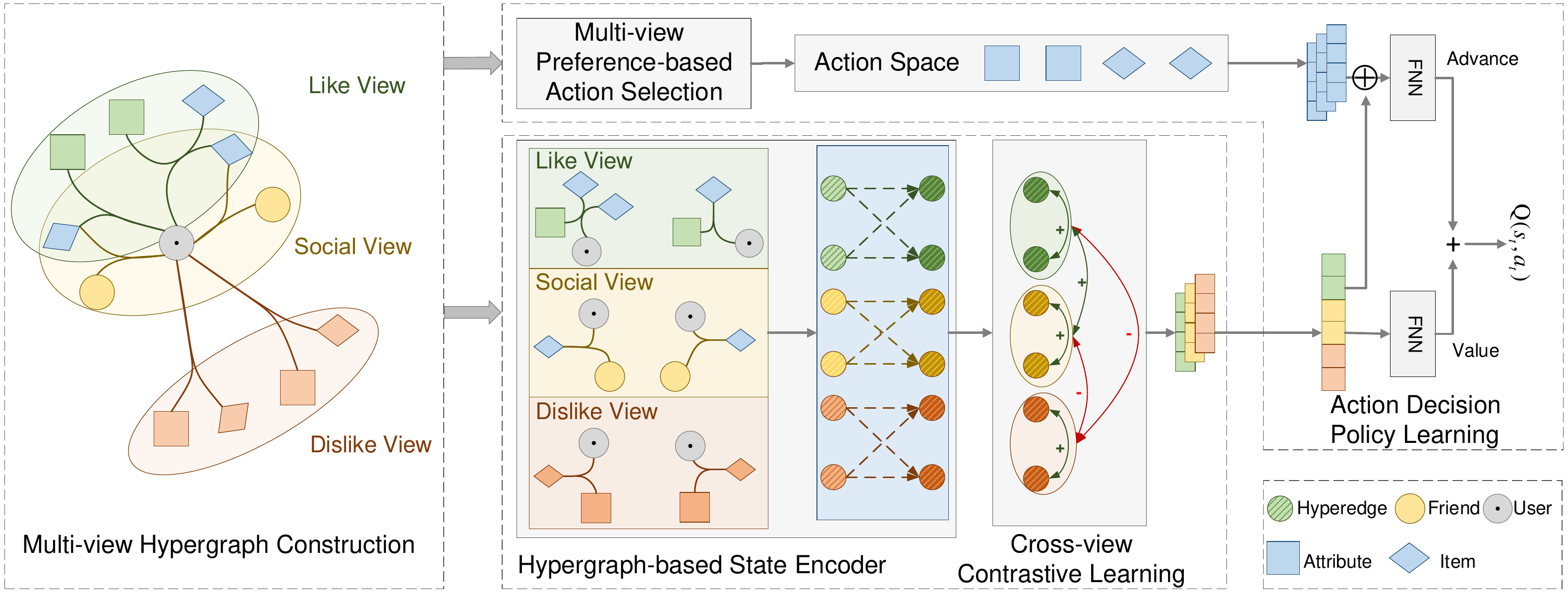}
    \caption{The overview of Multi-view Hypergraph Contrastive Policy Learning. It mainly contains four modules: (a) Multi-view Hypergraph Construction, which dynamically captures multiplex relations from three views. (b) Hypergraph-based State Encoder, which captures the graph structure and sequential modeling in the dynamic hypergraph. (c) Cross-view Contrastive Learning, which maintains the inherent characteristics and correlations of user preferences from different views, and (d) Action Decision Policy Learning to decide the next action based on the learned dynamic user preferences. (Best view in color.)}
    \label{fig:model}
\end{figure*}
As illustrated in Figure \ref{fig:model}, we model the user preference at timestep $t$ with a multi-view dynamic hypergraph which can be  formulated as $\mathcal{G}^{(t)}_u=(\mathcal{N}^{(t)}, \mathcal{H}^{(t)}, \mathbf{A}^{(t)})$, including: (1) a node set $\mathcal{N}^{(t)}=\{u\}\cup\mathcal{P}^{(t)}_{rej}\cup\mathcal{P}^{(t)}_{acc}\cup\mathcal{F}^{(t)}_{u}\cup\mathcal{V}_{p_0}$, where $\mathcal{V}_{p_0}$ indicates the items satisfying the initial attribute $p_0$ of the user $u$, and $\mathcal{F}^{(t)}_{u}$ denotes the filtered friends that have preferring items that satisfy the interactive history $\mathcal{I}^{(t)}_u$; (2) a hyperedge set $\mathcal{H}^{(t)}=\mathcal{H}^{(t)}_{like}\cup\mathcal{H}^{(t)}_{dis}\cup\mathcal{H}^{(t)}_{f}$,  where $\mathcal{H}^{(t)}_{like}$ denotes the user like items that satisfy the attribute (\emph{Like View}), $\mathcal{H}^{(t)}_{dis}$ indicates the user dislike items that satisfy the attribute (\emph{Dislike View}), and $\mathcal{H}^{(t)}_{f}$ denotes the user shares preferences to the items with the friend (\emph{Social View}). Each hyperedge $h \in \mathcal{H}^{(t)}$ is corresponding to an attribute $p_h$ or friend $f_h$; (3) a $|\mathcal{N}^{(t)}|\times |\mathcal{H}^{(t)}|$ adjacent matrix $\mathbf{A}^{(t)}$ which denotes the weighted edge between each node and hyperedge, with entries denoted as:
    \begin{equation}
    A_{i, j}^{(t)}= \begin{cases}
    1, & \text { if } n_{i}=u, h_j \in \mathcal{H}^{(t)}_{like} \cup \mathcal{H}^{(t)}_{f} \\
    -1, & \text { if } n_{i}=u, h_j \in \mathcal{H}^{(t)}_{dis} \\
    \frac{1}{|\mathcal{V}^{(t)}_{h_j}|}, & \text { if } n_{i}\in \mathcal{V}^{(t)}_{h_j}, h_{j}\in \mathcal{H}^{(t)}_{like} \cup \mathcal{H}^{(t)}_{dis} \\
    \frac{1}{|\mathcal{V}^{(t)}_{h_j}|}, & \text { if } n_{i}\in \mathcal{V}^{(t)}_{h_j}, h_{j}\in \mathcal{H}^{(t)}_{f} \\
    1, & \text { if } h_{j}\in \mathcal{H}^{(t)}_{like} \cup \mathcal{H}^{(t)}_{dis}, n_{i}= p_{h_j} \\ 
    1, & \text { if } h_{j}\in \mathcal{H}^{(t)}_{f}, n_{i}= f_{h_j} \\
    0, & \text { otherwise }\end{cases},
    \label{eq:hyperedge}
    \end{equation}
where $\mathcal{V}^{(t)}_{h_j}$ denotes items connected to the hyperedge $h_j$. Specifically, when $h_j \in \mathcal{H}^{(t)}_{like} \cup \mathcal{H}^{(t)}_{dis}$, $\mathcal{V}^{(t)}_{h_j}$ means items that satisfy the corresponding attribute $p_{h_j}$. And when $h_j \in \mathcal{H}^{(t)}_{f}$, it means the friend $f_{h_j}$'s acceptable items that satisfy the interactive history $\mathcal{I}^{(t)}_u$. We filter out noises in friends' acceptable items with the interactive history to help learn the user's current dynamic preferences.

\subsection{Hypergraph-based State Encoder}
\subsubsection{Hypergraph Message Passing Paradigm}
Motivated by the strength of hypergraph \cite{feng2019hypergraph, xia2022hypergraph} for generalizing the concept of edge to connect more than two nodes, we endow our MHCPL to capture multiplex relations under a hypergraph message passing architecture, where the hyperedges are treated as intermediate hubs for message passing across different nodes without the hop limitation. The formal representation of our hypergraph message passing is formulated as:
\begin{equation}
    \mathbf{\Gamma}=\text{ReLU}(\mathbf{A} \cdot \mathbf{H})=\text{ReLU}\left(\mathbf{A} \cdot \mathbf{A}^{\top} \cdot \mathbf{E}\right),
    \label{eq:hyperagg1}
    \end{equation}
where $\mathbf{E} \in \mathbb{R}^{|\mathcal{N}^{(t)}|\times d}$ denotes the initial embedding of nodes $\mathcal{N}^{(t)}$ in the hypergraph, $\mathbf{H} \in \mathbb{R}^{|\mathcal{H}^{(t)}|\times d}$ indicates the hyperedge representations aggregated from the node representations, and $\text{ReLU}(\cdot)$ denotes the LeakyReLU mapping. $\mathbf{\Gamma}$ denotes the hyper embedding of the nodes in the hypergraph representation space. With the help of hypergraph message passing, our MHCPL is capable to capture the multiplex collaborative relations that specify the attribute/friend that motivates/discourages the user's interest in the items.

\subsubsection{Hierarchical Hypergraph State Encoder} During the conversation, the hyperedges are successively generated when the user accepts or rejects the asked attribute. Moreover, the higher-level interactions between different hyperedges are also important in learning user preferences. Although the aforementioned hypergraph message passing paradigm is capable to capture the multiplex relations, it fails to model sequential information and hyperedge-wise feature interactions.  Inspired by the success of the Transformer encoder \cite{vaswani2017attention} in capturing sequential information and feature interactions, we employ the Transformer encoder to realize high-level hyperedge-wise message passing. Specifically, with the representation of hyperedges $\boldsymbol{H}$ that aggregate information from neighbor nodes, higher-level hypergraph layers further pass messages through the interactions between hyperedges under the same view:

%To deal with this problem, we further enhance our hypergraph neural architectures with higher-level hyperedge-wise feature interactions. Specifically, with the representation of hyperedges $\boldsymbol{H}$ that aggregate information from neighbor nodes, higher-level hypergraph layers further pass messages through the interactions between hyperedges under the same view as:
\begin{equation}
\begin{aligned}
    \bar{\mathbf{H}}=\psi^l(\boldsymbol{H}), \boldsymbol{H}=\mathbf{A}^{\top} \mathbf{E},
    \end{aligned}
    \label{eq:hyperagg2}
    \end{equation}
where $\mathbf{E} \in \mathbb{R}^{|\mathcal{N}^{(t)}|\times d}$ denotes the initial embedding of nodes $\mathcal{N}^{(t)}$ in the hypergraph. $\psi^l(\cdot)$ indicates the high-level hypergraph layers. $l$ denotes the layer number of high-level hypergraph layers. Hyperedges $\mathcal{H}^{(t)}_{o}$ of the same view $o \in \{like, dis, f\}$ are successively connected according to their generation order in the interactive conversation. To realize this, we apply the Transformer encoder ${MHSA}_o(\cdot)$ on hyperedges $\mathcal{H}^{(t)}_{o}$ of each view $o$ as:
\begin{equation}
\begin{aligned}
    \psi(\boldsymbol{H}^l_o)=\text{MHSA}_o(\boldsymbol{H}^{l-1}_{o}, \boldsymbol{H}^{l-1}_{o}, \boldsymbol{H}^{l-1}_{o}).
    \end{aligned}
    \label{eq:hyperagg3}
    \end{equation}
After the high-level hyperedge message passing, we aggregate the information from hyperedges to refine the node representations as:
\begin{equation}
\begin{aligned}
    \mathbf{\Gamma}_l=\text{ReLU}(\mathbf{A} \cdot \bar{\mathbf{H}})=\text{ReLU}\left(\mathcal{H} \cdot \psi^l\left(\mathbf{A}^{\top} \cdot \mathbf{E}\right)\right),
    \end{aligned}
    \label{eq:hyperagg4}
    \end{equation}
where $\psi^l$ denotes $l$ high-level hypergraph layers.
The hyper representation from different layers of the user is summed to get the representation of state $s_t$:
\begin{equation}
    \mathbf{q}_t=\sum_{l}\mathbf{\Gamma}_l(u)
\label{eq:hyperagg5}
\end{equation}
%In order to incorporate the hyper-connectivity information and sequential information into the representation of the state $s_t$, we employ transformer \cite{vaswani2017attention} and obtain the representation of state $s_t$ as:
%\begin{equation}
%\begin{aligned}
%    f_{\theta_{S}}\left(s_{t}\right)=\boldsymbol{N}^l_u\oplus\text{Mean}(\boldsymbol{H}^*),
%    \end{aligned}
%    \label{eq:hyperstate}
%    \end{equation}
%where $\boldsymbol{H}^{l}_{acc} \in \mathbb{R}^{|\mathcal{P}^{(t)}_{acc}|\times d}$ denotes the representation of the hyperedges corresponding to the accepted attribute instances in the historical interactions between the user and the CRS agent. $MHSA(\cdot)$ denotes the multi-head attention in transformer and $Mean(\cdot)$ denotes the mean pooing over the sequence. The presentation of the state $f_{\theta_{S}}\left(s_{t}\right)$ is obtained by concatenating $\oplus$ the representation of the user $\boldsymbol{N}^l_u$ and the interactive sequence.

\subsection{Cross-view Contrastive Learning}
Different types of multiplex relations present user preferences from various views (\ie \emph{Like View}, \emph{Dislike View}, \emph{Social View}). Actually, it is still non-trivial to sufficiently integrate user preferences from different views, since it might obscure the inherent characteristics of preference distributions from different views and the correlation between them.  Specifically, the user preferences from the same view should be more similar than those from different views. Also, user preferences from \emph{Like View} should be similar to \emph{Social View} while different from \emph{Dislike View}.
To capture these two correlations and better integrate user  preferences from different views, we develop cross-view contrastive learning based on InfoNCE\cite{oord2018representation} as:  
\begin{equation}
\begin{array}{ll}
    \mathcal{L}^{SSL}= \\
    -\sum\limits_{o}\sum\limits_{i \in \mathcal{H}_o }\log \frac{\sum\limits_{i^+ \in \mathcal{H}_o }\text{exp}({\operatorname{s}\left({{\mathbf{H}_i}}, {{\mathbf{H}_{i^+}} }\right) / \tau })}{ \underbrace{\sum\limits_{i^+ \in \mathcal{H}_o }\text{exp}({ \operatorname{s}\left({ {\mathbf{H}_i}}, {{\mathbf{H}_{i^+}}}\right) / \tau })}_{\text {positive pairs }} +
    \underbrace{\sum_{i^- \in \mathcal{H}-\mathcal{H}_o}\text{exp}({\operatorname{s}\left({{\mathbf{H}_i}}, {{\mathbf{H}_{i^-}}} \right)/ \tau }) }_{\text {negative pairs }}}\\
    -\sum\limits_{o}\sum\limits_{i \in \mathcal{H}_o }\log \frac{\sum\limits_{i^+ \in \mathcal{H}_{o^+} }\text{exp}({\operatorname{s}\left({{\mathbf{H}_i}}, {{\mathbf{H}_{i^+}} }\right) / \tau })}{ \underbrace{\sum\limits_{i^+ \in \mathcal{H}_{o^+} }\text{exp}({ \operatorname{s}\left({ {\mathbf{H}_i}}, {{\mathbf{H}_{i^+}}}\right) / \tau })}_{\text {positive pairs }} +
    \underbrace{\sum_{i^- \in \mathcal{H}_{o^-}}\text{exp}({\operatorname{s}\left({{\mathbf{H}_i}}, {{\mathbf{H}_{i^-}}} \right)/ \tau }) }_{\text {negative pairs }}}, 
    \end{array}
\label{eq:loss}
\end{equation}
where $o \in \{like, dis, f\}$ denotes three views, $\mathcal{H}=\mathcal{H}_{like}\cup\mathcal{H}_{dis}\cup\mathcal{H}_{f}$ indicates the set of hyperedges, $\mathbf{H}$ denotes the representations of hyperedges, and $s(\cdot)$ is the cosine similarity function. In Eq.\ref{eq:loss}, the first term is designed to maintain the intrinsic characteristics of user preferences from each view, which treats the hyperedges of the same view as positive pairs, while the different-view hyperedges as negative pairs. The second term of Eq.\ref{eq:loss} is designed to maintain the correlation of user preferences from different views, where the hyperedges in $\mathcal{H}_{like}$ and $\mathcal{H}_{f}$ are treated as positive pairs to each other, while treated as negative pairs with the hyperedges in $\mathcal{H}_{dis}$.

\subsection{Action Decision Policy Learning}
A large action search space reduces the efficiency of policy learning. Following \cite{deng2021unified,zhang2022multiple}, we select top-$\mathcal{K}_v$ candidate items and top-$\mathcal{K}_p$ candidate attributes to form the action space $\mathcal{A}_t$. To this end, we develop a multi-view action selection strategy, which selects items/attributes according to user preferences from three views. Specifically, we rank them as:
\begin{equation}
w_v^{(t)}=\sigma\left(\mathbf{e}_u^T \mathbf{e}_v+\sum_{p \in \mathcal{P}_{acc}^{(t)}} \mathbf{e}_v^T \mathbf{e}_{p}+\sum_{f \in \mathcal{F}_u^{(t)}} \mathbf{e}_v^T \mathbf{\tilde{e}}_f
-\sum_{p \in \mathcal{P}_{rej}^{\prime(t)}} \mathbf{e}_v^T \mathbf{e}_{p}\right),
\label{eq:itemscore}
\end{equation}

\begin{equation}
w_p^{(t)}=\sigma\left(\mathbf{e}_u^T \mathbf{e}_p+\sum_{p^{\prime} \in \mathcal{P}_{acc}^{(t)}} \mathbf{e}_p^T \mathbf{e}_{p^{\prime}}+\sum_{f \in \mathcal{F}_u^{(t)}} \mathbf{e}_p^T \mathbf{\tilde{e}}_f
-\sum_{p^{\prime} \in \mathcal{P}_{rej}^{\prime(t)}} \mathbf{e}_p^T \mathbf{e}_{p^{\prime}}\right),
\label{eq:attrscore}
\end{equation}
where $\mathbf{e}_u$, $\mathbf{e}_v$, $\mathbf{e}_p$ and $\mathbf{e}_f$ are embeddings of the user, item, attribute, and friend. $\mathbf{\tilde{e}}_f=\sum_{v^{\prime} \in \mathcal{V}_{f}^{(t)}} \mathbf{e}_v$ represents friend preferences that satisfy the interactive history, $\sigma(\cdot)$ denotes the sigmoid function.

With the action space $\mathcal{A}_t$ and the state representation $\mathbf{q}_t$, we introduce the dueling Q-networks \cite{wang2016dueling} to determine the next action and calculate the Q-value as:
\begin{equation}
Q\left(s_{t}, a_{t}\right)=f_{\theta_{V}}\left(\mathbf{q}_{t}\right)+f_{\theta_{A}}\left(\mathbf{q}_{t}, a_{t}\right),
\label{eq:DQN}
\end{equation}
where the value function $f_{\theta_{V}}(\cdot)$ and the advantage function $f_{\theta_{A}}(\cdot)$ are two separate multi-layer perceptions with $\theta_{V}$ and $\theta_{A}$ denote the parameters, respectively. The optimal Q-function $Q^{*}(\cdot)$, which has the maximum expected reward achievable by the optimal policy $\pi^*$, follows the Bellman equation \cite{bellman1957role} as:
\begin{equation}
Q^{*}\left(s_{t}, a_{t}\right)=\mathbb{E}_{s_{t+1}}\left[r_{t}+\gamma \max _{a_{t+1} \in \mathcal{A}_{t+1}} Q^{*}\left(s_{t+1}, a_{t+1} \mid s_{t}, a_{t}\right)\right],
\label{eq:DQN2}
\end{equation}
where $\gamma$ denotes the discount factor for the delayed rewards.

In each turn, the CRS agent will get the reward $r_t$, and we can update the state $s_{t+1}$ and the action space $\mathcal{A}_{t+1}$ according to the user's feedback. Following Deng \etal \shortcite{deng2021unified}, we define a replay buffer $\mathcal{D}$ to store the experience $\left(s_{t}, a_{t}, r_{t}, s_{t+1}, \mathcal{A}_{t+1}\right)$. For training of the DQN, we sample mini-batches from the buffer and minimize the following loss:
\begin{equation}
\mathcal{L}^{DQN}=\mathbb{E}_{\left(s_{a}, a_{t}, r_{t}, s_{t+1}, \mathcal{A}_{t+1}\right) \sim \mathcal{D}}\left[\left(y_{t}-Q\left(s_{t}, a_{t} ; \theta_{Q}, \theta_{S}\right)\right)^{2}\right],
\label{eq:DQN3}
\end{equation}
where $\theta_{S}$ is the set of parameters in the module for hypergraph-based representation learning, $\theta_{Q}=\left\{\theta_{V}, \theta_{A}\right\}$, and $y_{t}$ is the target value based on the currently optimal $Q^{*}$:

\begin{equation}
y_{t}=r_{t}+\gamma \max _{a_{t+1} \in \mathcal{A}_{t+1}} Q\left(s_{t+1}, a_{t+1} ; \theta_{Q}, \theta_{S}\right).
\label{eq:DQN4}
\end{equation}
To deal with the overestimation bias in the original DQN, we apply the double DQN \cite{van2016deep}, which copies a target network $Q^{'}$ as a periodic from the online network to train the model. During training, the action decision policy learning in Eq.\ref{eq:DQN3}, and the cross-view contrastive learning in Eq.\ref{eq:loss} are alternatively
optimize.

\section{Experiments}
% In this section, we will detail the settings of our experiments and present the experimental results.
To fully demonstrate the superiority of MHCPL,
we conduct experiments\footnote{https://github.com/Snnzhao/MHCPL} on two public datasets to explore the following questions:
\begin{itemize}
    \item \textbf{RQ1:} How does MHCPL perform compared with the state-of-the-art methods?
    \item \textbf{RQ2:} How do different components (social influence, hypergraph based state encoder, and cross-view contrastive learning) affect the results of MHCPL?
    \item \textbf{RQ3:} How do parameters (the layer number of Hypergraph based State Encoder) influence the results of MHCPL?
    \item \textbf{RQ4:} Can our MHCPL effectively leverage the interactive conversation, item knowledge, and social influence to learn the dynamic user preferences?
\end{itemize}

\subsection{Datasets}\label{sec:standalone}
To evaluate the proposed method, we adapt two existing
MCR benchmark datasets, named Yelp and LastFM. The statistics of these datasets are presented in Table \ref{tab:data}.
\begin{itemize}
    \item {\textbf{LastFM}}~\cite{lei2020estimation}: LastFM dataset is the music listening dataset collected from Last.fm online music systems. As Zhang \etal \shortcite{zhang2022multiple}, We define the 33 coarse-grained groups as attribute types for the 8,438 attributes.
    \item{\textbf{Yelp}}~\cite{lei2020estimation}: Yelp dataset is adopted from the 2018 edition of the Yelp challenge. Following Zhang \etal \shortcite{zhang2022multiple}, we define the 29 first-layer categories as attribute types, and 590 second-layer categories as attributes.
\end{itemize}
Following Zhang \etal \shortcite{zhang2022multiple}, we sample two items with partially overlapped attributes as the user's acceptable items for each conversation episode.
\begin{table}[ht]
    \setlength{\tabcolsep}{5pt}
 \centering
 \small
    \begin{tabularx}{0.45\textwidth}{p{3cm}|X|X}
    \toprule
    \makecell[c]{\text{Dataset}}&\makecell[c]{\text{Yelp}}&\makecell[c]{\text{LastFM}}\cr
    \hline
    \hline
    \makecell[c]{\text{Users}}&\makecell[c]{27,675}&\makecell[c]{1,801}\cr

    \makecell[c]{\text{Items}}&\makecell[c]{70,311}&\makecell[c]{7,432}\cr
    \makecell[c]{\text{Attributes}}&\makecell[c]{590}&\makecell[c]{8,438}\cr
    \makecell[c]{\text{Attribute types}}&\makecell[c]{29}& \makecell[c]{34}\cr
    \hline
    \makecell[c]{\text{User-Item}}&\makecell[c]{1,368,606}&\makecell[c]{76,693}\cr
    \makecell[c]{\text{User-User}}&\makecell[c]{688,209}&\makecell[c]{23,958}\cr
    \makecell[c]{\text{Item-Attribute}}&\makecell[c]{477,012}&\makecell[c]{94,446}\cr
    \bottomrule[0.8pt]
    \end{tabularx}
    \caption{Statistics of two utilized datasets}
    \label{tab:data}
\end{table}
\subsection{Experiments Setup}

\subsubsection{User Simulator}
MMCR is a system that is trained and evaluated based on interactive conversations with users. Following the user simulator adopted in \cite{zhang2022multiple}, we simulate a interactive session for each user-item set interaction pair $(u, \mathcal{V}_u)$. Each item in the item set $v \in \mathcal{V}_u$ is treated as an acceptable item for the user. Each session is initialized with a user $u$ specifying an attribute $p_0 \in \mathcal{P}_{joint}$, where $\mathcal{P}_{joint}$ is the set of attributes that are shared by the items in $\mathcal{V}_u$. Then the session follows the process of "System Ask or Recommend, User response" \cite{zhang2022multiple} as described in Section \ref{sec:def}.

\begin{table*}[t]
    \centering
    \begin{tabular}{p{2.0cm}<{\centering}p{1cm}<{\centering}p{1cm}<{\centering}p{1cm}<{\centering}p{1cm}<{\centering}p{1cm}<{\centering}p{0.01cm}p{1cm}<{\centering}p{1cm}<{\centering}p{1cm}<{\centering}p{1cm}<{\centering}p{1cm}<{\centering}p{0.01cm}p{1cm}<{\centering}p{1cm}<{\centering}p{1cm}<{\centering}p{0.01cm}p{1cm}<{\centering}p{1cm}<{\centering}p{1cm}<{\centering}}
    \toprule
    \multirow{2}{*}{\bfseries Models }&\multicolumn{5}{c}{\bfseries Yelp }&&\multicolumn{5}{c}{\bfseries LastFM }\\
    \cline{2-6}
    \cline{8-12}
    &SR@5&SR@10&SR@15&AT&hDCG&&SR@5&SR@10&SR@15&AT&hDCG\\
    \midrule

    Abs Greedy &0.078&0.124&0.150&13.65&0.065&&0.292&0.436&0.512&10.10&0.237\\
    Max Entropy&0.046&0.200&0.390&12.97&0.117&&0.280&0.560&0.680&9.34&0.263\\
    CRM& 0.026&0.100&0.188&13.99&0.059&&0.092&0.240&0.372&12.56&0.130\\
    EAR& 0.120&0.198&0.240&12.91&0.094&&0.298&0.436&0.508&10.08&0.237\\
    SCPR &0.146&0.188&0.436&12.29&0.169&&0.322&0.630&0.764&8.47&0.322\\
    UNICORN &\underline{0.200}&0.338&0.430&11.33&0.175&&0.444&0.774&0.846&7.10&0.348\\
     MCMIPL&{0.162}&{0.366}&{0.522}&{11.25}&{0.184}&&\underline{0.448}&{0.809}&{0.884}&{6.87}&{0.353}\\
    \midrule
    S*-UNICORN &0.120&0.478&0.696&10.59&0.223&&0.412&0.850&0.912&6.69&0.363\\
     S*-MCMIPL&{0.126}&\underline{0.490}&\underline{0.722}&\underline{10.51}&\underline{0.230}&&{0.442}&\underline{0.872}&\underline{0.940}&\underline{6.43}&\underline{0.368}\\
    \midrule
    MHCPL&{0.142}&{\bfseries 0.592}&{\bfseries 0.854}&{\bfseries 9.96}&{\bfseries 0.261}&&{\bfseries 0.470}&{\bfseries 0.938}&{\bfseries 0.982}&{\bfseries 5.87}&{\bfseries 0.427}\\
    % \midrule
     Improv. &-&20.82$\%$&18.28$\%$&5.23$\%$&13.48$\%$&&4.91$\%$&7.57$\%$&4.47$\%$&8.71$\%$&16.03$\%$\\
    \bottomrule
    \end{tabular}
    \caption{Performance comparison of different models on the two datasets. The bold number represents the improvement of our model over baselines is statistically significant with p-value $< 0.01$. hDCG stands for hDCG@($15,10$).}
    \label{tab:results}
\end{table*}

\subsubsection{Baselines}
To demonstrate the effectiveness of the proposed MHCPL, the state-of-the-art methods are chosen for comparison : 
\begin{itemize}
    \item \textbf{Max Entropy.} This method employs a rule-based strategy to ask and recommend. It chooses to select an attribute with maximum entropy based on the current state, or recommends the top-ranked item with certain probabilities \cite{lei2020estimation}.
    \item \textbf{Greedy\cite{christakopoulou2016towards}.} This method only makes item recommendations and updates the model based on the feedback. It keeps recommending items until the successful recommendation is made or the pre-defined round is reached. 

    \item \textbf{CRM\cite{sun2018conversational}.} A reinforcement learning-based method that records the users' preferences into a belief tracker and learns the policy deciding when and which attributes to ask based on the belief tracker.
    \item \textbf{EAR\cite{lei2020estimation}.} This method proposes a three-stage solution to enhance the interaction between the conversational component and the recommendation component.
    \item \textbf{SCPR\cite{lei2020interactive}.} This method learns user preferences by reasoning the path on the user-item-attribute graph via the user’s feedback and accordingly chooses actions.
    \item \textbf{UNICORN\cite{deng2021unified}.} This work builds a weighted graph to model dynamic relationships between the user and the candidate action space, and proposes a graph-based Markov Decision Process (MDP) environment to learn dynamic user preferences and chooses actions from the candidate action space.
    \item \textbf{MCMIPL\cite{zhang2022multiple}.} This approach proposes a multi-interest policy learning framework that captures the multiple interests of the user to decide the next action.
    \item \textbf{S*-UNICORN and S*-MCMIPL.} For a more comprehensive and fair performance comparison, we adapt UNICORN and MCMIPL by timely selecting helpful social information and incorporating it into the weighted graph of the model. We name the two adapted methods S*-UNICORN and S*-MCMIPL.
\end{itemize}

\subsubsection{Parameters Setting}
Following \cite{zhang2022multiple}, we recommend top $K=10$ items or ask $K_a= 2$ attributes in each turn. We employ the Adam optimizer with a learning rate of $1e-4$. Discount factor $\gamma$ is set to be $0.999$. Following \cite{deng2021unified}, we adopt TransE \cite{TransE} via OpenKE \cite{OpenKE} to pretrain the node embeddings with 64 dimensions in the constructed KG with the training set. We make use of Nvidia Titan RTX graphics cards equipped with AMD r9-5900x CPU (32GB Memory).
For the action space, we select $K_p=10$ attributes and $K_v=10$ items.
To maintain a fair comparison, we adopt the same reward settings as previous works  \cite{lei2020estimation, lei2020interactive,deng2021unified,zhang2022multiple}: $r_{rec\_suc}=1, r_{rec\_fail}=-0.1, r_{ask\_suc}=0.01, r_{ask\_fail}=0.1, r_{quit}=-0.3$. For MHCPL, we select the number of layers from {1, 2, 3, 4}.

\subsubsection{Evaluation Metrics}
Following previous works \cite{lei2020estimation, lei2020interactive,deng2021unified}, we adopt success rate (SR@t) to measure the cumulative ratio of successful recommendations by the turn t, average turns (AT) to evaluate the average number of turns for all sessions, and hDCG@(T, K) to additionally evaluate the ranking performance of recommendations. 
Therefore, the higher SR@t and hDCG@(T, K) indicate better performance, while the lower AT means an higher efficiency.

\subsection{Performance Comparison (RQ1)}
\subsubsection{Overall Performance}
The comparison experimental results of the baseline models and our models are shown in Table \ref{tab:results}. 
We can summarize our observations as follows:

\begin{itemize}[leftmargin=*]
    %our对比正常
    \item \textbf{Our proposed MHCPL achieves the best performance.} MHCPL significantly outperforms all the baselines on the metrics of SR@15, AT and hDCG by over 4.47$\%$, 5.23$\%$ and 13.48$\%$, respectively.  We attribute the improvements to the following reasons: 1) The proposed dynamic multi-view hypergraph could effectively capture multiplex relations from three views. And the proposed hierarchical hypergraph neural network is able to well learn dynamic user preferences by integrating the information of graph structure and sequential modeling from the dynamic multi-view hypergraph; 
    2) MHCPL timely selects helpful social information and effectively integrates the interactive conversation, item knowledge, and social influence for better dynamic user preference learning; 3) MHCPL designs a cross-view contrastive learning method to help maintain the inherent characteristics and the correlations of user preferences from different views.
    
    \item \textbf{The learning of the dynamic user preferences is crucial for conversational recommendation.} The graph-based methods (MHCPL, MCMIPL, UNICORN, SCPR) outperforms the factorization-based methods (EAR, CRM) since they learn user preferences from the collaborative information in the graph. MCMIPL achieves the best performance among the graph-base baselines since it further considers the multiple interests of the user preferences. Our proposed MHCPL further outperforms these methods since we leverage multiplex relations to integrate interactive conversation, item knowledge, and social influence to help learn the dynamic user preferences.

    \item \textbf{Social influence is effective in helping learn dynamic user preferences for conversational recommendation when well filtered.} The socially adapted methods (\ie S*-UNICORN and S*-MCMIPL) outperform their original versions in the final performances. We attribute this to the reason that social influence is an important factor that affects user preferences and could help learn dynamic user preferences with friends' preferences that satisfy the interactive conversation. But the socially adapted methods perform worse than their original version in the early turns (\eg SR@5). This happens because the information in the interactive conversation is not sufficient to filter out the noise from the social information in the early turn of the conversation.
    
\end{itemize}

\begin{figure}[t]
    \centering
    \begin{subfigure}{0.45\linewidth}
        \includegraphics[width=\textwidth]{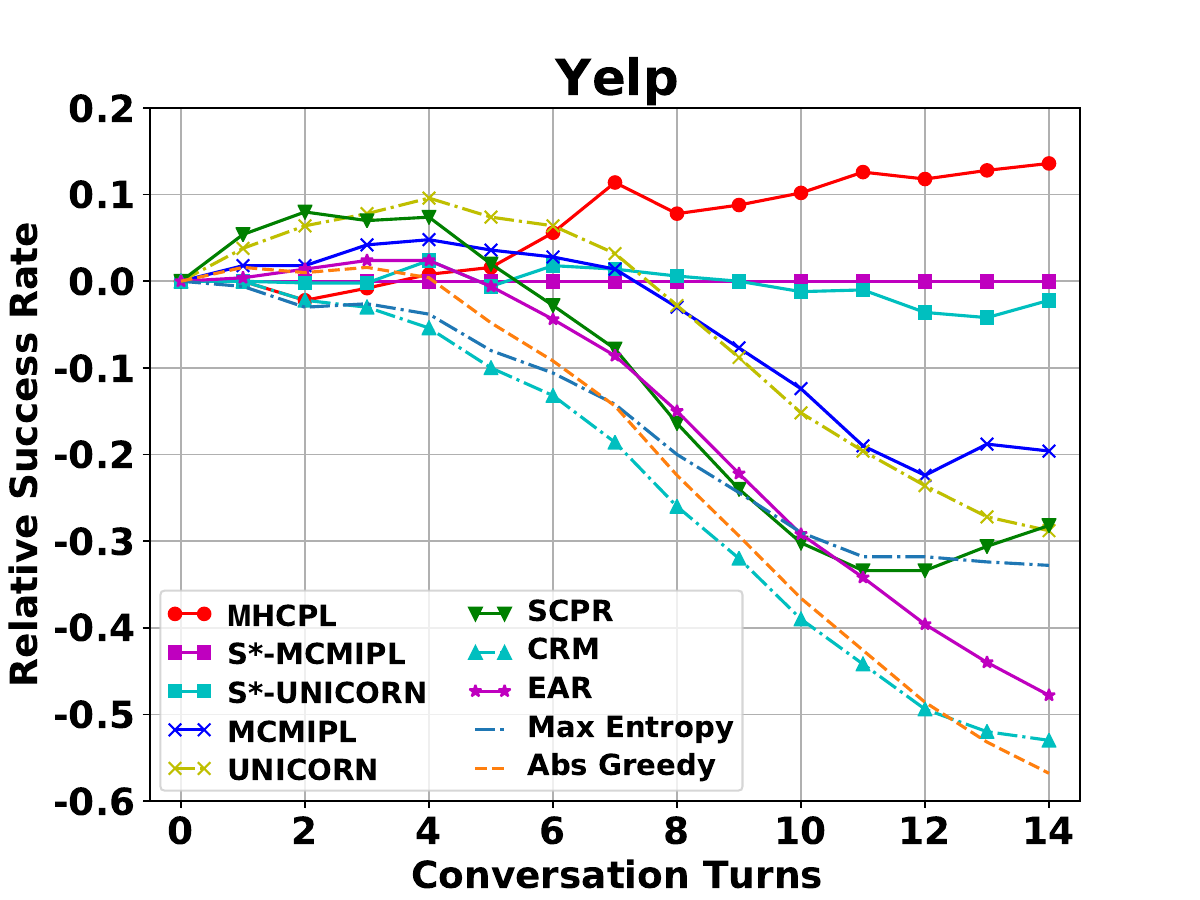}
        % \caption{}\label{fig:11}
    \end{subfigure}
    \begin{subfigure}{0.45\linewidth}
        \includegraphics[width=\textwidth]{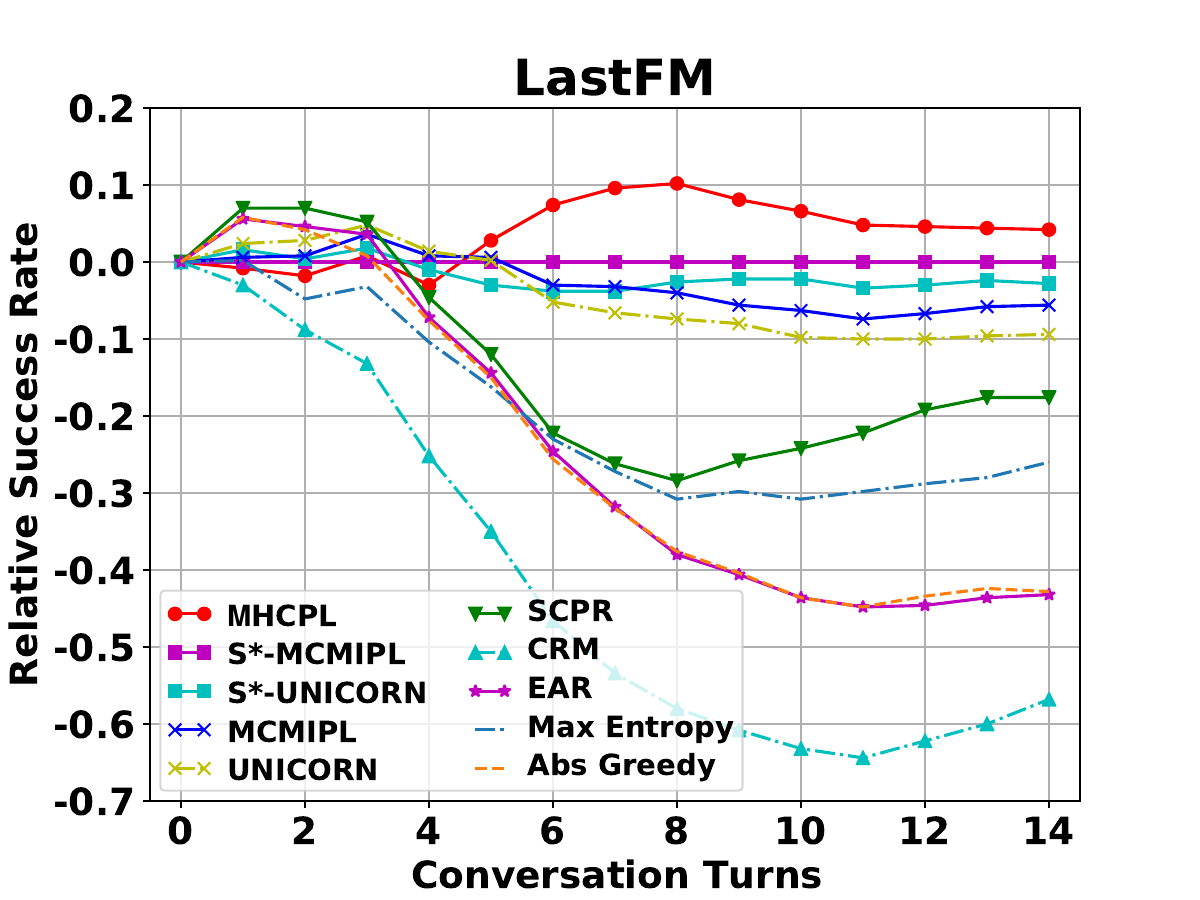}
        % \caption{}\label{fig:12}
    \end{subfigure}
    \caption{Comparisons at Different Conversation Turns.}
    \label{fig:overall}
\end{figure}

\subsubsection{Comparison at Different Conversation Turns} Besides the performance in the final turn, we also present success rates at different turns in \autoref{fig:overall}.
In order to better observe the differences among different models, we use the relative success rate compared with the most competitive baseline $\text{S*-MCMIPL}$, where the blue line of $\text{S*-MCMIPL}$ is set to zero in the figures. From the \autoref{fig:overall}, we following observations:
\begin{itemize}[leftmargin=*]
    %our对比正achieve
    \item \textbf{} The proposed MHCPL outperforms these baseline methods across all the datasets and almost all the turns in the conversational recommendation. This is because our proposed MHCPL could better learn dynamic user preferences with multiplex relations that integrate interactive conversation, item knowledge, and social influence.
    
    \item \textbf{} The recommendation success rate of the proposed socially-aware methods (\ie MHCPL, S*-MCMIPL, and S*-UNICORN) could not surpass all the baselines in the early turns of the conversational recommendation, especially on the dataset Yelp with a larger candidate space of items and attributes. This is because the information in the interactive conversation is not sufficient to filter out the noise from the social information at the early turn of the conversation.
    Furthermore, socially-aware methods prefer to ask rather than recommend in the early turns when the user's preference is not certain enough. This will effectively reduce the action space and better learn user preferences, but lead to a lower recommendation success rate in the early turns. %This is because they may recommend items with unclear user preferences in the early turns. This can increase early turns' success rates but is not effective since it is helpless in learning user preferences.
\end{itemize}
\begin{table}[t]
    \small
    \centering
    \begin{tabular}{p{2.7cm}<{\centering}p{0.6cm}<{\centering}p{0.5cm}<{\centering}p{0.6cm}<{\centering}p{0.6cm}<{\centering}p{0.5cm}<{\centering}p{0.6cm}<{\centering}}
    \toprule
    % \hline
    \multirow{2}{*}{\bfseries Models }& \multicolumn{3}{c}{\textbf{Yelp}}& \multicolumn{3}{c}{\textbf{LastFM}}\\
    &SR@15&AT&hDCG&SR@15&AT&hDCG\\
    \midrule
    Ours&{\bfseries0.854}&{\bfseries9.96}&{\bfseries0.261}&{\bfseries0.982}&{\bfseries5.87}&{\bfseries0.427}\\
    \midrule
    -w/o social&0.592&10.80&0.208&0.908&6.63&0.365\\
    -w/o hypergraph&0.726&10.68&0.346&0.938&6.58&0.382\\
    -w/o contrastive&0.762&10.37&0.237&0.962&6.17&0.403\\
    \bottomrule
    \end{tabular}
    \caption{Results of the Ablation Study. }
    \label{tab:ablation_study}
\end{table}
\subsection{Ablation Studies (RQ2)}
To investigate the underline mechanism of MHCPL, we conduct ablation experiments on the Yelp and LastFM datasets with three ablated methods including: $\text{MHCPL}_{\wo social}$ that ablates the social influence, $\text{MHCPL}_{\wo hypergraph}$ that replaces the hypergraph neural networks with graph neural networks, and $\text{MHCPL}_{\wo contrastive}$ that ablates the cross-view contrastive learning. From results shown in Table~\ref{tab:ablation_study}, we have the following observations:
\begin{itemize}
    \item $\text{MHCPL}_{\wo social}$ is the least  competitive. This demonstrates the importance of social influence in alleviating the data sparsity problem and helping learn dynamic user preferences. And it is effective to accordingly choose helpful social information based on interactive conversation. $\text{MHCPL}_{\wo social}$ still outperforms all the baselines that ignore the social information in Table~\ref{tab:results}, which proves the effectiveness of MHCPL in learning dynamic user preferences with multiplex relations.
    
    \item MHCPL outperforms $\text{MHCPL}_{\wo hypergraph}$. 
    We contribute this to the importance of multiplex relations in learning dynamic user preferences. This also proves the effectiveness of our proposed multi-view hypergraph-based state encoder in learning user preferences by integrating the information of graph structure and sequential modeling from the dynamic multi-view hypergraph.
    \item MHCPL outperforms $\text{MHCPL}_{\wo contrastive}$. This demonstrates the effectiveness of the cross-view contrastive learning module in helping maintain the inherent characteristics and correlations of user preferences from different views. 
\end{itemize}

\begin{figure}[!ht]
    \includegraphics[width=0.45\textwidth]{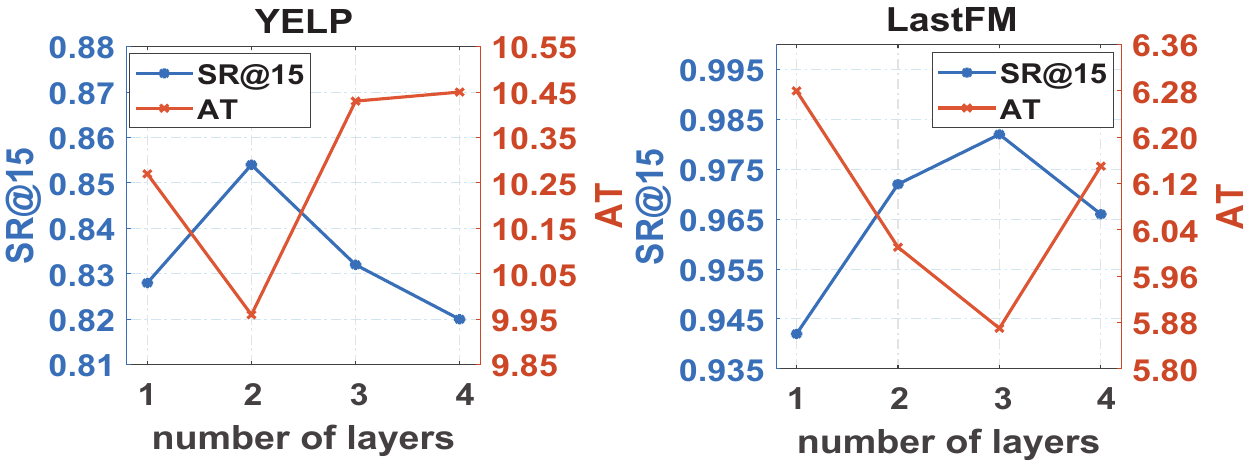}
    \caption{Impact of Layer Number(L)}
    \label{fig:layers}
  \end{figure}

\subsection{Hyper-parameter Sensitivity Analysis (RQ3)}
\subsubsection{Impact of Layer Number} The hypergraph-based state encoder learns dynamic user preferences from the multiplex relations in the hypergraph. By stacking more layers, collaborative information from multi-hop neighbors is distilled. We investigate how the layer number $L$ influences the performance of MHCPL. Specifically, we conduct experiments with $L$ in the range $\{1, 2, 3, 4\}$, and the results are shown in Figure \ref{fig:layers}. There are some observations:
\begin{itemize}
    \item Increasing the number of layers can improve the performance of our model. MHCPL-2 highly outperforms MHCPL-1. The reason is that MHCPL-1 only gains information from the one-hop neighbors and neglects high-order collaborative information. 
    \item When increasing the layer of number, the performance does not always improve. MHCPL-3 outperforms MHCPL-4 on data LastFM. This can be attributed to the noise which increases along with the hop of neighbors.
\end{itemize}

\begin{figure}[!ht]
    \includegraphics[width=0.45\textwidth]{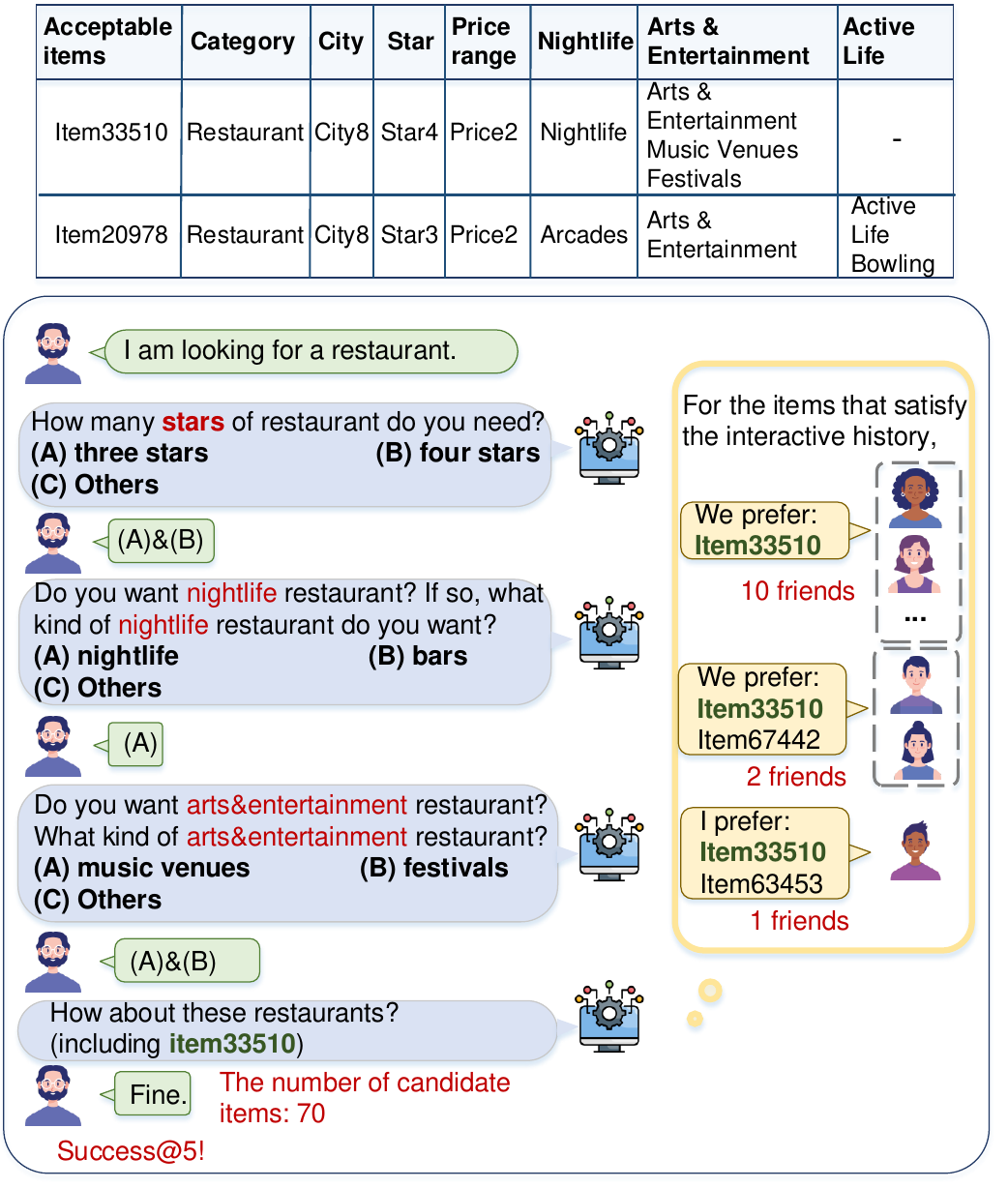}
    \caption{A case of conversation recommendation generated by our proposed MHCPL.}
    \label{fig:case}
  \end{figure}

\subsection{Case Study (RQ4)}
To show the effectiveness of our proposed MHCPL in leveraging multiplex relations to integrate interactive conversation, item knowledge, and social influence to learn dynamic user preferences, we present a case of conversational recommendation generated by our framework in \autoref{fig:case}. As illustrated in the figure, by integrating the information from the interactive conversation, item knowledge, and social information with multiplex relations from different views, MHCPL is able to effectively ask attributes and recommend user-preferred items, reaching success in five turns. Furthermore, the social information selected according to the interactive conversation is helpful in learning dynamic user preferences. With the help of selected social information, MHCPL could accurately select the target item when the information from the interactive history is limited in distinguishing user preferences towards the seventy candidate items.
\section{Conclusion}
In this work, we explore multiplex relations to integrate interactive conversation, item knowledge, and social influence in helping learn the dynamic user preferences for conversational recommendation. We propose a novel hypergraph-based model, namely \underline{M}ulti-view \underline{H}ypergraph \underline{C}ontrastive \underline{P}olicy \underline{L}earning (MHCPL), which timely selects useful social information according to the interactive history and builds a dynamic hypergraph with three types of multiplex relations from different views. 
A hierarchical hypergraph neural network is proposed to learn user preferences by integrating information of the graph structure and sequential modeling from the dynamic multi-view hypergraph.
Furthermore, a cross-view contrastive learning module is proposed with two terms to maintain the inherent characteristics and the correlations of user preferences from different views. 
Extensive experiments on two popular benchmarks demonstrate the superiority of our proposed method, as compared to the state-of-the-art baselines.

\section*{Acknowledgments}
This work was supported in part by the National Natural Science Foundation of China under Grant No.62276110, Grant No.61772076, in part by CCF-AFSG Research Fund under Grant No.RF20210005, and in part by the fund of Joint Laboratory of HUST and Pingan Property \& Casualty Research (HPL). The authors would also like to thank the anonymous reviewers for their comments on improving the quality of this paper. 
%This work was supported in part by the National Natural Science Foundation of China under Grant No.61602197, Grant No.L1924068, Grant No.61772076, in part by CCF-AFSG Research Fund under Grant No.RF20210005, and in part by the fund of Joint Laboratory of HUST and Pingan Property \& Casualty Research (HPL). The authors would also like to thank the anonymous reviewers for their comments on improving the quality of this paper.

%\normalem

\bibliographystyle{ACM-Reference-Format}
\bibliography{sigir2023}

\end{document}